\documentclass{ws-ijmpa}

\begin{document}

\markboth{B.A. Arbuzov, A.I. Pitikov and I.V. Zaitsev}
{Non-perturbative approach to anomalous FB asymmetry in top pair production}

%
%
\title{Non-perturbative approach to the problem of anomalous FB asymmetry in top pair production at TEVATRON}
\author{Boris A. Arbuzov}
\address{\it Skobeltsyn Institute of Nuclear Physics of
MSU,\\119991 Moscow, Russia\\
arbuzov@theory.sinp.msu.ru}
\author{Alexei I. Pitikov}
\address{\it Department of Physics of
MSU,\\119991 Moscow, Russia\\
}
\author{Ivan V. Zaitsev}
\address{\it Skobeltsyn Institute of Nuclear Physics of
MSU,\\119991 Moscow, Russia\\
zaitsev@theory.sinp.msu.ru}

\maketitle

\begin{history}
\received{Day Month Year}
\revised{Day Month Year}
\end{history}

\begin{abstract}
We apply Bogoliubov compensation principle to the problem of forward-backward asymmetry in $t \bar t$ production at TEVATRON. The non-trivial solution of compensation equation for anomalous four-fermion interaction effective interaction of doublets of heavy and light quarks leads to possibility of existence of $\bar t u,\, \bar b d$ vector bound states (resonance). The coupling of the bound state to $\bar t,\,u$ pair is calculated. With this result we obtain satisfactory description of totality of data on the effect of FB asymmetry including slope of effect dependence on invariant $t \bar t$ mass. Predictions for the resonance mass and production cross-section at LHC are presented..

\keywords{compensation equation; effective four-fermion interaction; forward-backward asymmetry.}
\end{abstract}

\ccode{PACS numbers: 12.15.-y, 11.30.Qc, 11.15.Wx}

Recently results were presented on measurement of Forward-Backward asymmetry in $t \bar t$ production in $p \bar p$ collisions at energy 
$\sqrt{s}\,=\,1960\,$GeV~\cite{CDF,D0,CDF2}: $A_{FB} = 0.158\pm 0.075$~\cite{CDF}, $A_{FB} = 0.196\pm 0.065$~\cite{D0}, $A_{FB} = 0.162\pm0.047$~\cite{CDF2}. According to the results the asymmetry significantly exceeds the corresponding SM calculations $A_{FB} = 0.089\pm 0.007$~\cite{SM}.
  
In present work we make attempt to explain the anomaly in terms of non-perturbative effects in SM, considering spontaneous generation of non-local four-quark interaction. In previous works we have used Bogoliubov compensation method \cite{Bog1,Bog2}  as a mechanism for generation of non-local vertices, which produce non-perturbative effects in corresponding theories. Particularly, in work \cite{AZ1} we obtained Bogoliubov compensation equation solution for $t \bar t$ quarks' interaction, and made upon this ground some conclusions about composite Higgs properties. Below we shall start from this solution, but just for our present problem we have to take into consideration also mixed $\bar u t$ quarks' interaction. Accordingly, we have to investigate compensation equation for corresponding vertex and watch for influence, which it can produce in anomalous t-quark interactions.

Bogoliubov compensation principle (which quantum field theory adaptation in detail described in \cite{Arb04,Arb05,AVZ} allow us to obtain cutting form-factors for non-renormalizable  vertices as a solution  of special relations, expressing demand of mutual compensation of these vertices different order contributions into free quark Green functions. So, as a result we can use such free propagators in their usual form, and in the same time hold in our theory non-renormalizable  interactions without divergence in integrals.

Practically for mentioned relations construction we use "add-subtract" procedure: we insert additional term $\Delta L$ into Lagrangian both with "+" and "--" signs, and then one of this term we takes into free Lagrangian $L_0$, and other - into interaction Lagrangian $L_{int}$. So, we must then demand, that just the first addition together with constructed by this element higher order contributions into field propagation functions mutually compensate themselves. And the difference in sign in the same time will provide us preservation of analogous term in interaction Lagrangian: here compensation will not take place.

In our theory we take following additional term:
\begin{equation}
 \Delta{L}= \frac{\it{G_1}}{2}\bullet \,{\overline{\Psi}}^a_{L}{\gamma}_{\alpha}{\psi}^a_{L}\times{\overline{\psi}}^b_{L}{\gamma}_{\alpha}{\Psi}^b_{L} +\frac{\it{G}}{2}\bullet \,{\overline{\Psi}}^a_{L}{\gamma}_{\alpha}{\Psi}^a_{L}\times{\overline{\Psi}}^b_{L}{\gamma}_{\alpha}{\Psi}^b_{L}\,.\label{vertex}
\end{equation}
Here $\Psi$ and $\psi$ - correspondingly heavy quark doublet $(t,\,b$ and light quark doublet $(u,\,d$,$\;\Psi_{L}=\frac{1-\gamma_{5}}{2}\,\Psi\;$, $\psi_{L}=\frac{1-\gamma_{5}}{2}\,\psi$,$\newline\,a,b$ are the colour indexes and symbols like $\frac{\it{G}}{2}\bullet $ denotes non-local vertex, writing in momentum space as
\begin{equation}2\pi^{4} G\,\,\gamma_{\alpha}\times \gamma_{\alpha}\,F(p,q,k,r)\,;
\end{equation}
with p,q,k,r - incoming or outgoing momenta of each line and F - form-factor function.
Firstly we consider special kinematic conditions, setting $q=-p$,$\;k=r=0$ as it presented in Fig.1 and thus we have all diagrams depending only on one value $p^2$. Note, that color here is  transmitted along lines.
\begin{center}
\includegraphics[scale=0.6]{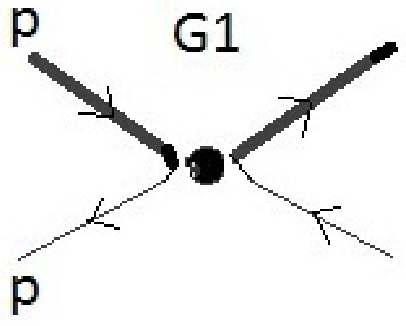}\\
Fig. 1. Diagram for vertex~(\ref{vertex}).

\end{center}
\bigskip     
The approximation, which we are working out here, proposes also following assumptions, where experience acquired in the course of previous works determines our choice:\\
1) In compensation equation we restrict ourselves by
terms with loop numbers 0, 1, 2.\\
2) We reduce thus obtained non-linear compensation equation to a linear
integral equation. It means that in loop terms only one vertex
contains the form-factor, being defined above, while
other vertices are considered to be point-like.\\
3) While evaluating diagrams with point-like vertices diverging
integrals appear. Bearing in mind that as a result of the study we
obtain form-factors decreasing at momentum infinity, we use an
intermediate regularization by introducing UV cut-off $\Lambda$ in the
diverging integrals. Final results will  not depend on the
value of this cut-off.\\
4) We shall not consider contribution to our equations of vertices of type ${\overline{\Psi}}^a_{L}\,\frac{\widehat{p}\,p_\rho}{p^2}\,{\Psi}^a_{L}\times{\overline{\Psi}}^b_{L}\,\frac{\widehat{p}\,p_\rho}{p^2}\,{\Psi}^b_{L}$. Thus we multiply the equation by projecting operator $({\gamma}_{\rho}-\frac{\widehat{p}\,p_\rho}{p^2})$, just eliminating such terms and not affecting our standard vertices (see also~\cite{AVZ2}).

These assumptions allow us to build compensation equation for vertex $G_1$, which in diagram form presented at Fig.2 (as it was mentioned above, such equation for vertex G we obtained in previous work).
\begin{center}
\includegraphics[scale=0.5]{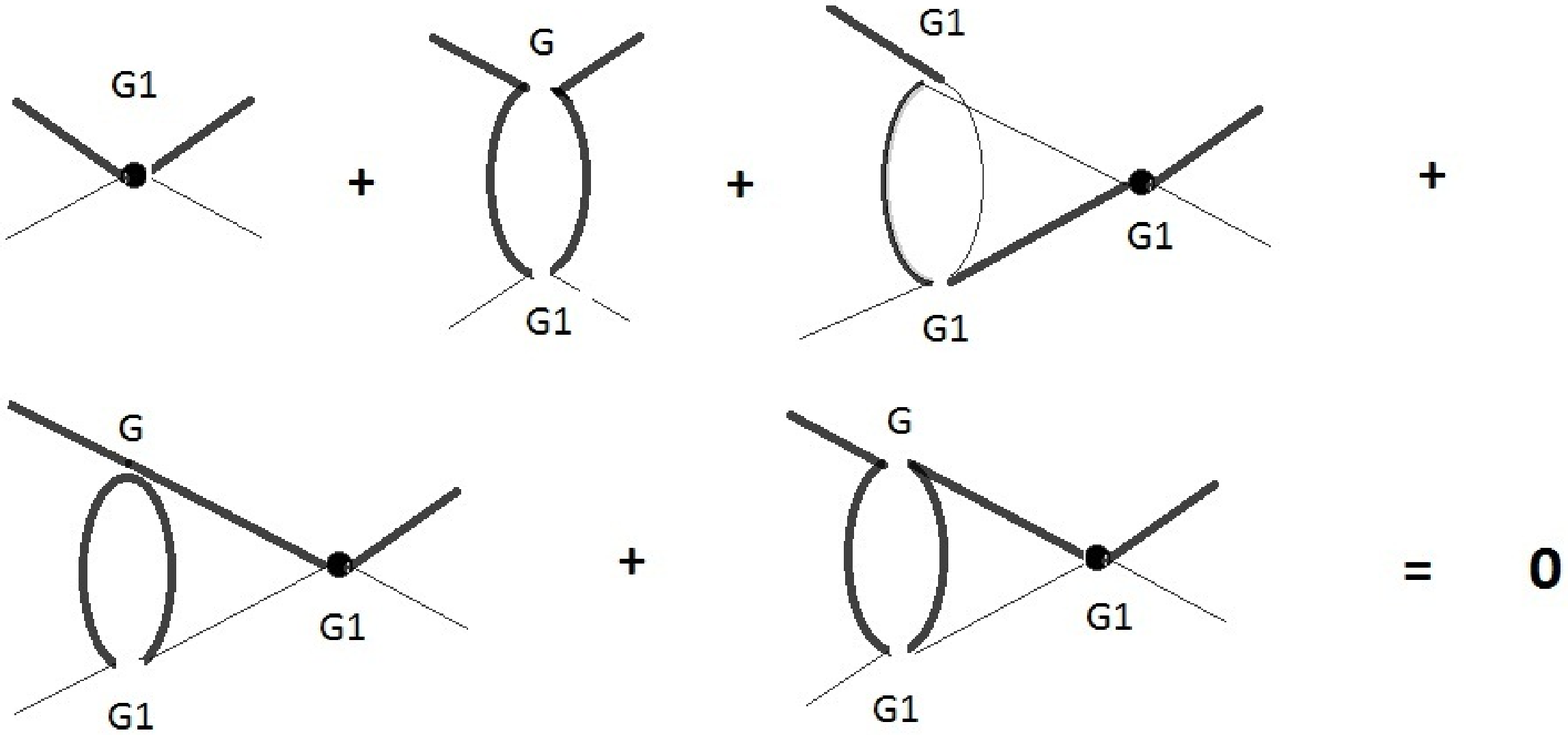}\\
\bigskip
Fig. 2. Diagram representation of compensation equation for 
form-factor of interaction~(\ref{vertex}). Here solid lines represents t-quark, thin ones - u-quark. Also solid point represents vertex with form-factor $F(p^2)$, whereas in the rest cases we have point-like four-fermion vertexes.
\end{center}

After angle integration we obtain following expression:
\begin{eqnarray}
& &F \left( x \right)\, +\,\frac{\it{G}}{16 {\pi}^2}\left(-11/18\,x+1/3\,x\ln  \left( {\frac {x}{{\Lambda}^{2}}} \right) +\,{\Lambda}^2\right)
\nonumber\\
& &
+{\frac {1}{3072\,{\pi }^{4}}}\, \Bigl(\left( 10\,{\it G_1}\,G-24\,{{\it G_1^{2}}} \right) \ln  \left( x \right) x \int _{0}^{x}\!F \left( y \right)  {dy}\nonumber\\
& &- 24\,{\it {G_1^{2}}}  \ln
 \left( x \right)\int _{0}^{x}\!F \left( y
 \right) y  \,{dy}+\left(\frac{14}{3} \,{\it G_1}\,G - 40\, {{\it G_1^{2}}}\right)\int \!\,F \left( y \right) y\, {dy}\nonumber\\
& &  -\left( 2\,{{\it G_1^{2}}}+{\frac {5}{2}}\,{\it G_1}\,G \right) \frac{1}{x}\,\int _{0}^{x}\!F \left( y \right){y}^{2} {dy} +\left( {\frac {7}{5}}\,{\it G_1}\,G -\frac{2}{5}\,{{\it G_1^{2}}}\right) \frac{1}{{x}^{2}}\int _{0}^{x}\!F \left( y \right) {y}^{3} {dy}\label{intEq}\nonumber\\
& &-24\,{
\it {G_1}}^{2} \int _{x}^{\infty }\!F \left( y \right)   \ln  \left( y \right) y\,{dy}+x \left( 10\,{\it G_1}\,G-24\,{{\it G_1^{2}}}
 \right)\int _{x}^{\infty }
\!F \left( y \right)  \ln  \left( y \right) {dy}\nonumber\\ & &-\left( 2\,{\it G_1}\,G+40\,{{\it G_1^{2}}} \right) x\int _{x}^{\infty }\!F \left( y
 \right)  {dy}+ \left( 5\,{\it G_1}
\,G-2\,{{\it G_1^{2}}} \right) {x}^{2}
\int _{x}^{\infty }\!{\frac {F \left( y \right) }{y}}{dy}\nonumber\\ & &+\left( {\frac {17}{30}}\,{\it G_1}\,G-\frac{2}{5}\,{{\it G_1^{2}}} \right) {x}^{3}\int _{x}^{\infty }\!\frac {F \left( y \right) }{y^{2}}{dy}+ \left( 36\,{{\it G_1^{2}}}-10\,{\it G_1}\,G \right)\int _{0}^{\infty }\!F \left( y \right)  y\,{dy}\nonumber\\ & &+ \left(  \left( -10\,{\it G_1}\,G
+24\,{{\it G_1^{2}}} \right) \ln  \left( {\Lambda}^{2} \right) +36\,{{
\it G_1^{2}}}-\frac{55}{3}\,{\it G_1}\,G \right) x\int _{0}^{\infty }\!F \left( y \right) {dy}\nonumber\\ & &+  24\,{{\it G_1^{2}}}\ln \left( {\Lambda}^{2} \right)  \int _{0}^{\infty }\!F
 \left( y \right)  y\,
 {dy}\nonumber+ \left( 30\,{\it G_1}\,G - 48\,{{\it G_1^{2}}}\right) {\Lambda}
^{2}\int _{0}^{\infty }\!F \left( y
 \right) {dy}
\biggr)=0\,.\nonumber
\end{eqnarray}
Here $x=p^2$, $y=q^2$ with q - loop integrating momenta variable, and $\Lambda$ - intervening momenta integration cut-off. Further results will not depend on this $\Lambda$. We count color number as $N=3$.

This expression  provides an equation of the type which were
studied in previous works ~\cite{Arb04,Arb05,AVZ},
where the way of obtaining solutions of such  equations was described. Indeed, by successive differentiation of this relation we obtain Meijer equation:
\begin{eqnarray}
& &\biggl(x\frac{d}{dx}\biggr)\biggl(x\frac{d}{dx}\biggr)\biggl(x\frac{d}{dx} -1\biggr)\biggl(x\,\frac{d}{dx}-1\biggr)\biggl(x\frac{d}{dx}- 2\biggr)\biggl(x\frac{d}{dx}-3\biggr)\biggl(x\frac{d}{dx}+1\biggr)\times\nonumber\\ 
& &\biggl(x\frac{d}{dx}+2\biggr)F(x)+{\frac {{{\it G_1^{2}}}{x}^{2} }{6144 \,x^2 \,{\pi }^{4}}}\,\left( 4\, \left( {\frac {d^{4}}{d{x}^{4}}}
F \left( x \right)  \right) \xi\,{x}^{4}+48\, \Biggl( {\frac {d^{3}}{d{
x}^{3}}}F \left( x \right)  \right) \xi\,{x}^{3}+\label{dif}\\ 
& & \left( {\frac {d
^{2}}{d{x}^{2}}}F \left( x \right)  \right)(144\, \xi -12) {x}^{2}+ \left( {\frac {d}{dx}}F \left( x \right)  \right) \left(81\,\xi -48\right)x+F \left(
x \right)\left(-30\,\xi+12\right)\Biggr)=0 .\nonumber 
\end{eqnarray}
Here we take
\begin{equation}
\it{G}=\xi\,\it{G_1}\,.
\end{equation}
Making substitution of the variable
\begin{equation}
z=\xi\,{\frac {{{\it G_1^{2}}} }{{1536\,\pi }^{4}}}{x}^{2}\,;
\end{equation}
we  obtain solution in terms of a Meijer function~\cite{be}
\begin{equation}
F\left(z\right)=\it{C}\,G_{48}^{52}\Bigl( z\,|^{a_1,\,a_2;\,a_3,\,a_4}_{3/2,\,1,\,1/2,\,1/2,\,0;\,0,\,-1/2,\,-1}\Bigr)\,;\label{solution}
\end{equation}
where parameters $a_1,\,a_2,\,a_3,\,a_4$ depends on $\xi$ and this choice of their distribution presuppose, that solution for $\xi$ will give us two first, "active" parameters with negative real part, and two last - with positive.

To determine $\xi$ and $C$ we consider two relation. First of them is  normalization condition for form-factor $F$:
\begin{equation}
F\left(0\right)=1;\label{norm}
\end{equation}
Then we can mention, that we have inhomogeneous term  with $x\rightarrow 0$ of order of $x\,ln\left(x\right)$. So, it has to be eliminated by analogous term in Meijer function expansion.

Near $z=0$ we have
\begin{equation}
\,G_{48}^{52}\Bigl( z\,|^{a_1,\,a_2;\,a_3,\,a_4}_{3/2,\,1,\,1/2,\,1/2,\,0;\,0,\,-1/2,\,-1}\Bigr)\sim 16/3\,{\frac {\Gamma  \left( 3/2-{ a_1} \right) \Gamma  \left( 3/2-{
 a_2} \right) }{\Gamma  \left( { a_3}-1/2 \right) \Gamma  \left(
{ a_4}-1/2 \right) }}\,\sqrt{z}\,ln\left(z\right)\,.
\end{equation}
Taking also proportional $x\,ln(x)$ term from the inhomogeneous part of the equation we obtain
\begin{equation}16/3\,{\frac {\,\Gamma  \left( 3/2-{ a_1} \right) \Gamma  \left( 3/2-{
 a_2} \right) }{\Gamma  \left( { a_3}-1/2 \right) \Gamma  \left(
{ a_4}-1/2 \right) }}\,C=\sqrt{\frac{\xi}{6}}
\end{equation}
Thus we obtain $\it{C}$, and substituting it into ~(\ref{norm}), we have
\begin{equation}\xi=1.33623\,;
\end{equation}
and, as a result,
\begin{equation}F\left(z\right)=0.56229\,G^{5 2}_{4 7}\left(z\, \Big\vert\,^{a_1,\,a_2,\,a_3,\,a_4}_{3/2,\, 1, 1/2, 1/2,\, 0,\, - 1/2,\, - 1}\right)\,;\end{equation}
where
\begin{eqnarray}
& & a_1\,=\,- 0.53192- \imath\,0.410037\,;\quad a_2\,=\,- 0.53192+ \imath\,0.410037\,;\nonumber\\
& & a_3\,=\,1.39470\,;\quad a_4\,=\,0.66913\,.\label{ai}
\end{eqnarray}
Since we have this solution for form-factor function, we can take into consideration bound state of interacting fields. Let us consider Bethe-Salpeter equation for a vector bound state. Its easy to see, that in our case this equation can be obtained by simple sign changing in front of the kernel of the compensation equation~(\ref{intEq}) (and also exclusion of inhomogeneous terms). So, it means, that we shall obtain as a solution Meijer function with the same parameters, as we just have in form-factor function, but with their changed distribution. Namely, we have such solution for normalized to 1 at zero momentum Bethe-Salpeter wave function $\Psi$
\begin{equation}\Psi\left(z\right)={\frac {\Gamma  \left( { a_3} \right) \Gamma  \left( {a_4}
 \right)
}{\Gamma  \left( 1-{a_1} \right) \Gamma  \left( 1-{ a_2} \right)
}}\,G^{4 2}_{4 8}\left(z\, \Big\vert\,^{{ a_1}, { a_2}, { a_3}, { a_4}}_{3/2, 1, 1/2, 0, 1/2, 0, -1/2, -1}\right)\,;
\end{equation}
with $a_i$~(\ref{ai}).  
And then we can write normalization condition for bound state propagator, demanding, that linear by squared incoming momentum term in expansion of presented below diagram  turn itself to $1$.
\bigskip  
\begin{center}
\includegraphics[scale=0.6]{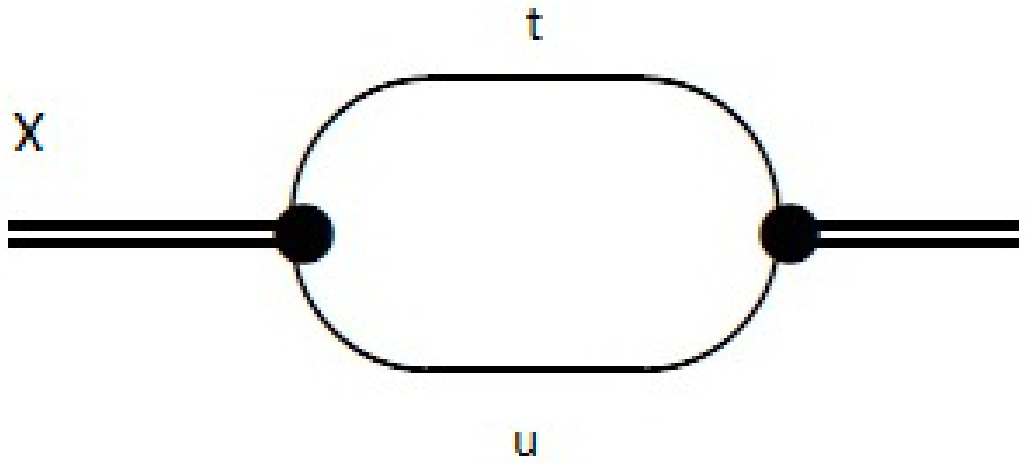}\\
Fig. 3. Diagram for normalization of the Bethe-Salpeter wave function.
\end{center}
\bigskip  
It gives value of coupling constant $g$ of this bound state, standing in the vertices of diagram:
\begin{equation}
g=\frac{\pi}{\sqrt{I_1}}\, ,\quad I_1=\frac{1}{2}\int _{\mu}^{\infty }\! \frac{\Psi^2\left(z\right)}{z}\,;\label{BSnorm}
\end{equation}
where we write $\mu$ at lower integral limit instead of entering  mass into propagator's denominator. We can estimate $\mu$ taking corresponding value $\mu_0$ from the results of heavy quark bound state investigation in work~\cite{AZ1} and accounting integration variable changing:
\begin{equation}
\mu_0=4.0675*10^{-12}\,,\quad \mu=\frac{3}{8\,\xi}\,\mu_0
\end{equation}
Thus we obtain following results:
\begin{equation}
I_1=12.077\, ,\quad g=0.904\,;
\end{equation}

So, using these results we can estimate physical effects under  discussion and compare obtained predictions with experimental data.

Forward-backward asymmetry in proton collisions defined in such way:
\begin{equation}
A_{FB}=\frac{N(cos(\theta) >0)-N(cos(\theta) < 0)}{N(cos(\theta) >0)+N(cos(\theta) < 0)}\,;
\end{equation}
where $\theta$ is the angle between proton and top-quark. CDF and D0 Collaboration results for $A_{FB}$ are correspondingly
\begin{equation}A_{FB}=0.162 \pm 0.047 \qquad \textrm{and} \qquad A_{FB}=0.196 \pm 0.065\,.
 \end{equation}

 We can also take into consideration invariant $(t\,\bar t)$ mass distribution of the asymmetry:
\begin{equation}
A_{FB}(M_{t\overline{t}})=\frac{N_{F}(M_{t\overline{t}})-N_{B}(M_{t\overline{t}})}{N_{F}(M_{t\overline{t}})+N_{B}(M_{t\overline{t}})},
\end{equation}
Figure below shows CDF data together with SM predictions excluding background:
\begin{center}
\includegraphics[scale=0.5]{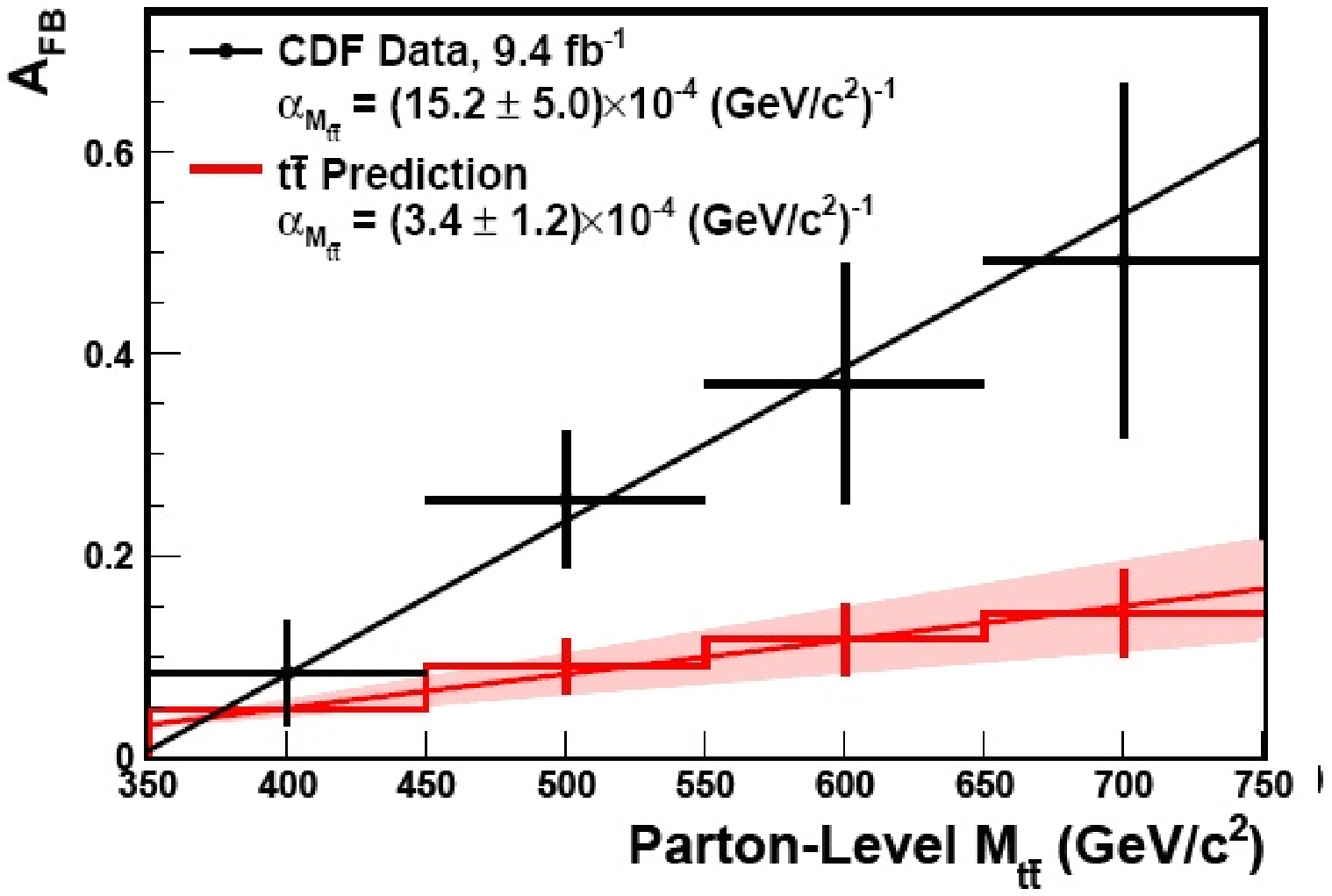}\\
Fig. 4. Invariant mass $M_{t \bar t}$ distribution of the asymmetry. Lower line presents SM calculations.  
\end{center}
At parton level we have the following SM prediction
 \begin{equation}
\alpha_{M_{t\overline{t}}}=(3.4\pm1.2)\times 10^{-4}(\textrm{GeV})^{-1} 
 \end{equation}
 and the experimental result~\cite{CDF2} 
 \begin{equation}
\alpha_{M_{t\overline{t}}}=(15.2\pm5.0)\times 10^{-4}(\textrm{GeV})^{-1}  
 \end{equation}

Calculations of the effects are done with applications of CompHEP system~\cite{comphep}
 
We add to SM interactions our additional interaction between introduced bound state X and quarks:
\begin{equation}
L_{fX}= g\, ( \,\bar \Psi\,\gamma_{\alpha}(1+\gamma_5){\psi}\times \bar X_{\alpha}+X_{\alpha}\times \bar {\psi}{\gamma}_{\alpha}(1+\gamma_5){\Psi})\,.\label{tuX}
\end{equation}
For three values of parameter g from interval $g=0.9\pm0.1$ we can demand, that total cross-section of $p\overline{p}\rightarrow t\overline{t}$ process reaches value of $\sigma=7.8\pm0.4 \,pb$ and thus obtain corresponding value of $M=M_X$. Then accounting all possible sub-processes we can estimate for each g  value of asymmetry A  as also slope $\alpha$. Results presented in Tables 1 and 2.
\bigskip
\begin{center}
Table 1.\\
FB asymmetry for different values of coupling $g$ and bound state X mass $M$ (GeV).
\end{center}
 \bigskip
\begin{center}
\begin{tabular}[l]{|||c|||c||c|||c||c|||c||c|||
} \hline
$\sigma$&\multicolumn{2}{|c|||}{$8.2 \, pb$}&\multicolumn{2}{c|||}{$7.8 \, pb$}&\multicolumn{2}{c|||}{$7.4 \, pb$}\\ \hline
 g&$M $&$A_{FB}\%   $&$M $&$A_{FB}\% $&$M $&$A_{FB}\% $\\
\hline 0.8&528&  $22.21\pm 0.07$ &550 & $18.4\pm 0.05$ &575 & $14.5\pm 0.04$ \\
 \hline 0.9&608& $21.4\pm0.06$
&632& $17.8\pm0.05$&660& $14.1\pm0.04$\\
 \hline 1.0&686& $20.7\pm0.06$&713& $17.3\pm0.04$
&747& $13.5\pm0.04$\\
 \hline \hline
 \end{tabular}
\end{center}
\bigskip

\begin{center}
\bigskip

Table 2.\\
Slope $\alpha$ in $10^{-4}/\textrm{GeV}$ for different values of coupling $g$ and bound state X mass $M$ in GeV.
\end{center}
 
\begin{center}
\begin{tabular}[l]{|||c|||c||c|||c||c|||c||c|||
} \hline
$\sigma$&\multicolumn{2}{|c|||}{$8.2 \, pb$}&\multicolumn{2}{c|||}{$7.8 \, pb$}&\multicolumn{2}{c|||}{$7.4 \, pb$}\\ \hline
 g&$M $&$\alpha  $&$M $&$\alpha $&$M $&$\alpha $\\
\hline 0.8&528&  $18.23 $ &550 & $17.14 $ &575 & $16.34 $ \\
 \hline 0.9&608& $19.91$
&632& $18.62$&660& $16.99$\\
 \hline 1.0&686& $19.41$&713& $17.33$
&747& $16.32$\\
 \hline \hline
 \end{tabular}
\end{center}

The results, presented in Tables, show, that our proposal can 
describe the existing data on the effect. Indeed the calculated values of asymmetry $A$ and of slope $\alpha$ are inside error bars of experimental results~\cite{CDF,D0,CDF2} for whole intervals of variation $0.8 < g < 1.0$ and $7.4\,pb < \sigma < 8.2\,pb$. 

It is interesting to compare with data from LHC. First of all the forward-backward asymmetry can not be measured at LHC due to symmetry of the initial state $ p\,p$. We also are to emphasize, that the dominant contribution to the $t \bar t$ production at LHC is given by gluon-gluon fusion, in which contribution of the mechanism under discussion is absent. However the production of the new states $X,\, \bar X$ can contribute to the total cross-section of $t \bar t$ production. The estimate of this contribution for $g = 0.9$ and $M = 632\,GeV$ gives the following value for $\sqrt{s} = 7\,TeV$
\begin{equation}
\Delta\,\sigma\,=\,11.6 \,pb\,;\label{LHCtt}
\end{equation} 
where we have taken into account value of the branching ratio 
\begin{equation}
BR(X \to t\,\bar u)\,=\,0.47\,;\quad \Gamma_X\,=\,154\,GeV\,.\label{BRXtu}
\end{equation} 
Value~(\ref{LHCtt}) is inside error bars of $t$-pair production cross-section at LHC~\cite{LHCTTATL,LHCTTCMS,LHCTTCMS12}. For example, 
results of works~\cite{LHCTTATL,LHCTTCMS12} read correspondingly
\begin{eqnarray}
& &\sigma(pp  \to t\,\bar t+...)\,=\,176 \pm 5\,^{+14}_{-11} \pm 8\,pb\,;\label{ttLHC}\\
& &\sigma(pp  \to t\,\bar t+...)\,=\,158 \pm 2.1\pm 10.2 \pm 3.5\,pb\,;\nonumber
\end{eqnarray} 
while the SM calculations give
\begin{equation}
\sigma(pp  \to t\,\bar t+...)\,=\,165^{+11}_{-16}\,pb\,.
\end{equation} 
We see, that possible additional contribution~(\ref{LHCtt}) causes no trouble yet.

However we are to emphasize, that our variant predicts the production of $X,\,\bar X$ resonances and it is quite advisable to look for this effect in the forthcoming LHC studies.
Let us remind parameters of these resonances: mass in the range  $600\, -\,700\,GeV$, width around $150\,GeV$ and production cross section for $\sqrt{s} = 7\,TeV$ in channel $X \to t\,\bar u$ is of the order of magnitude~(\ref{LHCtt}).
The resonance can be also observed in the channel 
$X \,\to\,b\,\bar d$ with branching ratio $0.53$. Note, that 
our approach also give results for processes with $b\,\bar b$ production, which are of the same order of magnitude as for t-quarks. However, it seems, that background conditions for 
$b\,\bar b$ processes are more severe.  

In case of confirmation of the present option we would get the result, that the asymmetry under discussion by no means contradict the Standard Model. On the other hand this would mean necessity to take into account non-perturbative effects in SM, in particular, by application of the Bogoliubov compensation approach.

\end{document}